\documentclass[aps, prd, letterpaper, 11pt, amsmath, amssymb, amsfonts, floatfix, nofootinbib, superscriptaddress]{revtex4-2}
\usepackage[utf8]{inputenc}

\usepackage{amsmath,amssymb}
\usepackage{slashed}
\usepackage{siunitx}
\usepackage{bm}
\usepackage{color}
\usepackage{graphicx} 
\usepackage[colorlinks=true,linkcolor=blue,citecolor=blue]{hyperref}
\usepackage{feynmp-auto}
\usepackage{dsfont}
\usepackage{cancel}
\usepackage{accents}
\usepackage{mciteplus,slashed}
\usepackage{amssymb,cancel,amsmath}
\usepackage{dcolumn}
\usepackage{bm}
\usepackage[caption=false]{subfig}
\usepackage{appendix}
\usepackage{feynmp-auto}
\usepackage[T1]{fontenc}	
\usepackage{csvsimple}

\usepackage[dvipsnames]{xcolor}
\usepackage[normalem]{ulem}

\usepackage{verbatim}

\usepackage{natbib}
\usepackage{cleveref}

\pdfoutput=1 
             \pdfoutput=1
\usepackage[sort&compress]{natbib}
\usepackage{url}
\usepackage{hyperref}
\usepackage{amsfonts}
\usepackage{epsfig}
\usepackage{cancel}
\usepackage{dcolumn}

\usepackage{amsmath}
\usepackage{amssymb}
\usepackage{amsthm}
\usepackage{amsfonts}
\usepackage{verbatim}
\usepackage{enumerate}
\usepackage{graphicx}
\usepackage{accents}
\usepackage{bm}

\usepackage{physics}
\usepackage{tensor}
\usepackage{siunitx}
\usepackage{circuitikz}
\usepackage{wasysym}

\usepackage{graphics}
\usepackage{caption}
\usepackage{epstopdf}
\usepackage[utf8]{inputenc} 
\usepackage{pgfplots}
\usepackage{slashed}

\definecolor{gr}{rgb}{0.0, 0.42, 0.24}

\begin{document}
\unitlength=1.3mm

\title{\boldmath The Smallest SU($N$) Hadrons}
\author{Stefano~Profumo}
\email{profumo@ucsc.edu}
\affiliation{Department of Physics, University of California Santa Cruz, 1156 High St., Santa Cruz, CA 95064, USA and \\
Santa Cruz Institute for Particle Physics, 1156 High St., Santa Cruz, CA 95064, USA}

\begin{abstract}
If new physics contains new, heavy strongly-interacting particles belonging to irreducible representations of SU(3) different from the adjoint or the (anti)fundamental, it is a non-trivial question to calculate what is the minimum number of quarks/antiquarks/gluons needed to form a color-singlet bound state (``hadron''){, or, perturbatively, to form a gauge-invariant operator, with the new particle}. Here, I prove that for an SU(3) irreducible representation with Dynkin label $(p,q)$, the minimal number of quarks needed to form a product that includes the (0,0) representation is $2p+q$.  I generalize this result to SU($N$), with $N>3$. {I also calculate the minimal total number of quarks/antiquarks/gluons that, bound to a new particle in the $(p,q)$ representation, give a color-singlet state, or, equivalently, the smallest-dimensional gauge-invariant  operator that includes quark/antiquark/gluon fields and the new strongly-interacting matter field. Finally, I list all possible values of the electric charge of the smallest hadrons containing the new exotic particles, and discuss constraints from asymptotic freedom both for QCD and for grand unification embeddings thereof.}
\end{abstract}

\maketitle
\flushbottom
\newpage 

\section{Introduction}
\label{sec:intro}
In Quantum Chromo-Dynamics (QCD), a gauge theory with gauge group SU(3) that describes the strong nuclear force in the Standard Model of particle physics, color confinement is the phenomenon that color-charged particles cannot be isolated, i.e. cannot subsist as stand-alone asymptotic states. From a group-theoretical standpoint, quarks belong to the fundamental representation of SU(3), antiquarks to the antifundamental representation, and the force mediators, gluons, to the adjoint representation\footnote{In what follows, I will both use the notation $\boldsymbol{d}$ to indicate a representation of dimension $d$ and $\boldsymbol{\overline{d}}$ to indicate the corresponding conjugate representation, and the notation $(p,q)$, with $p$ and $q$ non-negative integers; I use the convention that the bar corresponds to representations where $q>p$. The dimension of representation  $(p,q)$ is $d=(p+1)(q+1)(p+q+2)/2$. For instance, quarks belong to the irreducible representation $\boldsymbol{3}\sim(1,0)$, antiquarks to $\boldsymbol{\overline{3}}\sim(0,1)$, and gluons to $\boldsymbol{8}\sim(1,1)$ (for details see \cite{Slansky:1981yr,Dynkin:1957um}; for an exhaustive review on Lie groups see e.g. \cite{Yamatsu:2015npn})}.  
Color-confinement can thus be stated in group-theoretic language as the phenomenon that asymptotic, physical states must belong to the singlet (trivial) representation of SU(3), which I indicate below as $\boldsymbol{1}\sim(0,0)$. For instance, in real life physical states of strongly-interacting particles include mesons, which are quark-antiquark states, belonging to the singlet representation resulting from $\boldsymbol{3}\otimes\boldsymbol{\overline{3}}=\boldsymbol{8}\oplus\boldsymbol{1}$; and baryons, which are three-quark states, belonging to the singlet representation resulting from $\boldsymbol{3}\otimes\boldsymbol{3}\otimes\boldsymbol{3}=\boldsymbol{10}\oplus\boldsymbol{8}\oplus\boldsymbol{8}\oplus\boldsymbol{1}$. In addition, glueballs, bound states of gluons, could also exist \cite{MATHIEU_2009}, since $\boldsymbol{8}\otimes\boldsymbol{8}=\boldsymbol{27}\oplus\boldsymbol{10}\oplus\boldsymbol{\overline{10}}\oplus\boldsymbol{8}\oplus\boldsymbol{8}\oplus\boldsymbol{1}$.

Here, I am interested in which bound states would form around a hypothetical new particle $X$ charged under SU(3) and belonging to some irreducible representation of SU(3) with Dynkin label $(p,q)$. { In perturbative QCD, the answer to the same question also provides the form of the  smallest-dimensional gauge-invariant operator containing quark, antiquark, and gluon fields, and the new particle $X$}. Here, I specifically address two questions: the first, simple question is how many ``quarks'' would be needed to form a color-less bound state, i.e. what is the minimal number of copies of the fundamental representation such that the direct product of those copies and of the $(p,q)$ contains the trivial representation $(0,0)$. The answer is $2p+q$: I prove this in two different ways below. I then generalize this result to SU($N$). Secondly, I pose the slightly less trivial question of what is the minimal number of ``elementary constituents'', i.e. quarks, antiquarks and ``valence gluons'', needed to form a colorless bound state with the new particle $X$ {( or, again, in the language of perturbative QCD what is the smallest-dimensional gauge-invariant operator combining quark, antiquark, and gluon fields and the new matter field).}

{ While it is meaningless to ``count'' the number of gluons in a hadron in a non-perturbative sense, the notion of ``valence gluon'' plays an important role in a variety of contexts, including the study of the mass spectrum of bound states (usually dubbed $R$-hadrons) of gluinos in supersymmetry \cite{Chanowitz:1983ci, Buccella:1985cs}, in Yang-Mills-Higgs theories (see e.g. \cite{Oxman:2012ej}) and, more generally, the  phenomenology of new particles charged under SU(3) (see e.g. \cite{Chung:1997rz,Kachelriess:2003yy, Hewett:2004nw,  Kang:2006yd, Albuquerque:2009qy, DelNobile:2009st, delAguila:2010mx,Han:2010rf,Ilisie:2012cc,Kats:2012ym,Bertuzzo:2012bt, DeLuca:2018mzn,  Bramante_2019}). 

At least in the real world, the ``smallest'' hadrons (protons, neutrons, pions) are also the lightest ones in the spectrum, and there are good reasons to believe that the same could be true for a new exotic heavy state. Specifically, bag models \cite{DeGrand:1975cf,DeRujula:1975qlm,Rebbi:1975ns}, that include information on valence gluons, have been instrumental in estimating the mass spectrum of $R$-hadrons ever since the seminal work of Ref.~\cite{Chanowitz:1983ci} (see also \cite{Buccella:1985cs}); the multi-jet phenomenology of new strongly-interacting massive particles also depends on the properties of the product of representations containing multiple gluon fields, see e.g. \cite{Kumar:2011tj}. In this (perturbative) context, rather than the existence of a bound state, the key is which gauge-invariant operators exist containing the new strongly interacting state and a given number of quark, antiquark, and gluon fields. This informs, in turn, the multi-jet structure to be expected at high-energy colliders \cite{Kumar:2011tj}. 

From the standpoint of astro-particle physics, the existence of new, stable colored particles has been widely considered as well, see e.g. \cite{Chung:1997rz, Albuquerque:2009qy, Hewett:2004nw, Kachelriess:2003yy}; under some circumstances, such new colored states could even be the dark matter, or a part thereof (see e.g. \cite{Kang:2006yd,DeLuca:2018mzn,Bramante_2019}). Electric charge neutrality, however, restricts which irreducible representations the new strongly-interacting matter field can belong to. The questions I address here are thus relevant to several aspects of the associated phenomenology, such as whether the bound states are electrically charged, and which number of jets are expected from inelastic interactions with the parton fields in nucleons in the atmosphere, with implications for the shower structure at ground-based telescopes \cite{Chung:1997rz, Albuquerque:2009qy, Hewett:2004nw}.

Since limits on new strongly-interacting states imply that the mass of the $X$ be much higher than the QCD scale $\Lambda_{\rm QCD}$ \cite{Aaboud:2019trc,Aaboud:2018hdl,Rappoccio:2018qxp}, any state containing more than one $X$, such as for instance the color-singlet $\overline{X}X$, would be significantly heavier than any bound state of $X$ with quarks, antiquarks or gluons. Additionally, the absence of new strongly-interacting states at the Large Hadron Collider (LHC) implies that any such new state be generically heavier than the electroweak scale \cite{Aaboud:2019trc,Aaboud:2018hdl,Rappoccio:2018qxp}.

In what follows I consider the results for all SU(3) representations with dimension smaller than 100, including the minimal number of quark, antiquark, and gluons that combined with the $X$ provide an SU(3) singlet state, and the lowest-possible mass dimension of gauge-invariant operators involving the $X$. As a corollary, I calculate the possible values of the electric charge of the ``smallest hadron'' $H$ containing $X$, and I list all possible $(p,q)$ irreducible representation such that the ``smallest'' hadron can be electrically neutral. I also outline results for $N>3$, but leave the detailed exploration of the general case to future work}.

The reminder of the paper is organized as follows: in the next section \ref{sec:proof} I provide two proofs that the minimal product of fundamental representations of SU(3) is $2p+q$ and generalize the result to SU($N$); {in the following section \ref{sec:minimal} I calculate the composition of the ``smallest hadrons'' in SU(3), which, as mentioned, is equivalent to calculating the lowest-possible mass dimensional gauge-invariant operators containing the new strongly-interacting state and Standard Model fields, and outline the calculation for SU($N$), $N>3$; in sec.~\ref{sec:asfreedom} I discuss constraints from asymptotic freedom, and the embedding of the $X$ particle in a grand unification setup, also elaborating upon asymptotic freedom in the resulting grand unified theory; the final sec.~\ref{sec:conclusions} concludes.}

\section{The minimal direct product of fundamental representations of SU($N$) containing the trivial representation}\label{sec:proof}
Irreducible representations of SU($N$) are conveniently displayed with Young tableaux via the following rules (for more details, see e.g. \cite{Itzykson:1965hk, Lichtenberg:1978pc, Greiner:1989eu, Georgi:1999wka}):
\begin{enumerate}
    \item[(i)] The fundamental representation is represented by a single box;  
     \item[(ii)] Young tableaux for SU($N$) are left-justified $N-1$ rows of boxes such that any row is not longer than the row above it;
     \item[(iii)] Any column with $N$ boxes can be crossed out as it corresponds to the trivial (singlet) representation.
\end{enumerate}

Any irreducible representation can be obtained from direct products of the fundamental representation; the direct product of two representations proceeds via the following rules:
\begin{enumerate}
    \item[(i)] Label the rows of the second representation's tableau with indices $a,\ b,\ c,....$, e.g.\\
    
    \begin{figure}[!h]
        \centering
        \includegraphics[width=3cm]{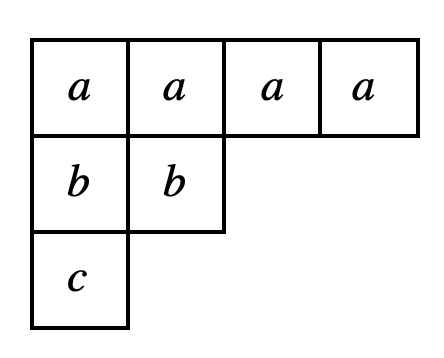}
    \end{figure}

     \item[(ii)] Attach all boxes from the second to the first tableau, one at a time, following the order $a,\ b,\ c,....$, in all possible way; the resulting Young tableaux is admissible if it obeys the rules above, and if there are no more than one $a,\ b,\ c,....$ in every column;
     \item[(iii)] Two tableaux with the same shape should be kept only if they have different labeling;
     \item[(iv)] A sequence of  indices $a,b,c,...$ is admissible if at any point in the sequence at least as many $a$’s have occurred as $b$’s, at least as many $b$’s have occurred as $c$’s, etc.; all tableaux with indices in any row, from right to left, arranged in a non-admissible sequence must be eliminated.
\end{enumerate}

The direct product of $k$ fundamentals is especially simple, since it entails repeated attachment of one additional box up to $k$ new boxes to any row, if that operation produces an admissible tableau (for instance, one cannot attach a box to a row containing as many boxes as the row above).

In the case of SU(3), Young tableaux have only two rows, and can be labeled with the Dynkin indices $(p,q)$, where $q$ is the number of boxes in the second row, and $p$ the number of additional boxes in the first row with respect to the second (thus, the first row has $p+q$ boxes). The dimensionality of the representation is given by
\begin{equation}
    {\rm dim}\ =\ \frac{1}{2}\left(p+1\right)\left(q+1\right)\left(p+q+2\right);
\end{equation}
similar formulae exist for $N>3$.

The direct product of the fundamental and a generic irreducible representation $(p,q)$ generally includes 
\begin{equation}\label{eq:fund}
    (p,q)\otimes (1,0)=(p+1,q)+(p-1,q+1)+(p,q-1)
\end{equation}
where the last two representations only exist if $p\ge1$ and $q\ge1$, respectively. As a result, to obtain the singlet representation $(0,0)$ from $(p,q)$ we need exactly $p$ copies of the fundamental to bring $p\to0$ (visually, by adding the extra boxes all to the second row); these will bring us to the representation $(0,q+p)$; at that point, we attach $q+p$ boxes to the third row (i.e. multiply by an additional $q+p$ fundamentals) to obtain the singlet representation.

The operational sequence outlined above is also the most economical, since, as Eq.~(\ref{eq:fund}) shows, $p$ can only decrease by one unit for each additional fundamental representation factor, but doing so costs an increment of one unit to $q$; similarly, $q$ can also only decrease by one unit at a time, thus {\em the minimal number $k$ of fundamental representations needed to obtain a representation that includes the singlet representation from the direct product of a given representation $(p,q)$ and $k$ copies of the fundamental representation is $k=2p+q$}.

\begin{figure}
    \centering
    \includegraphics[width=10cm]{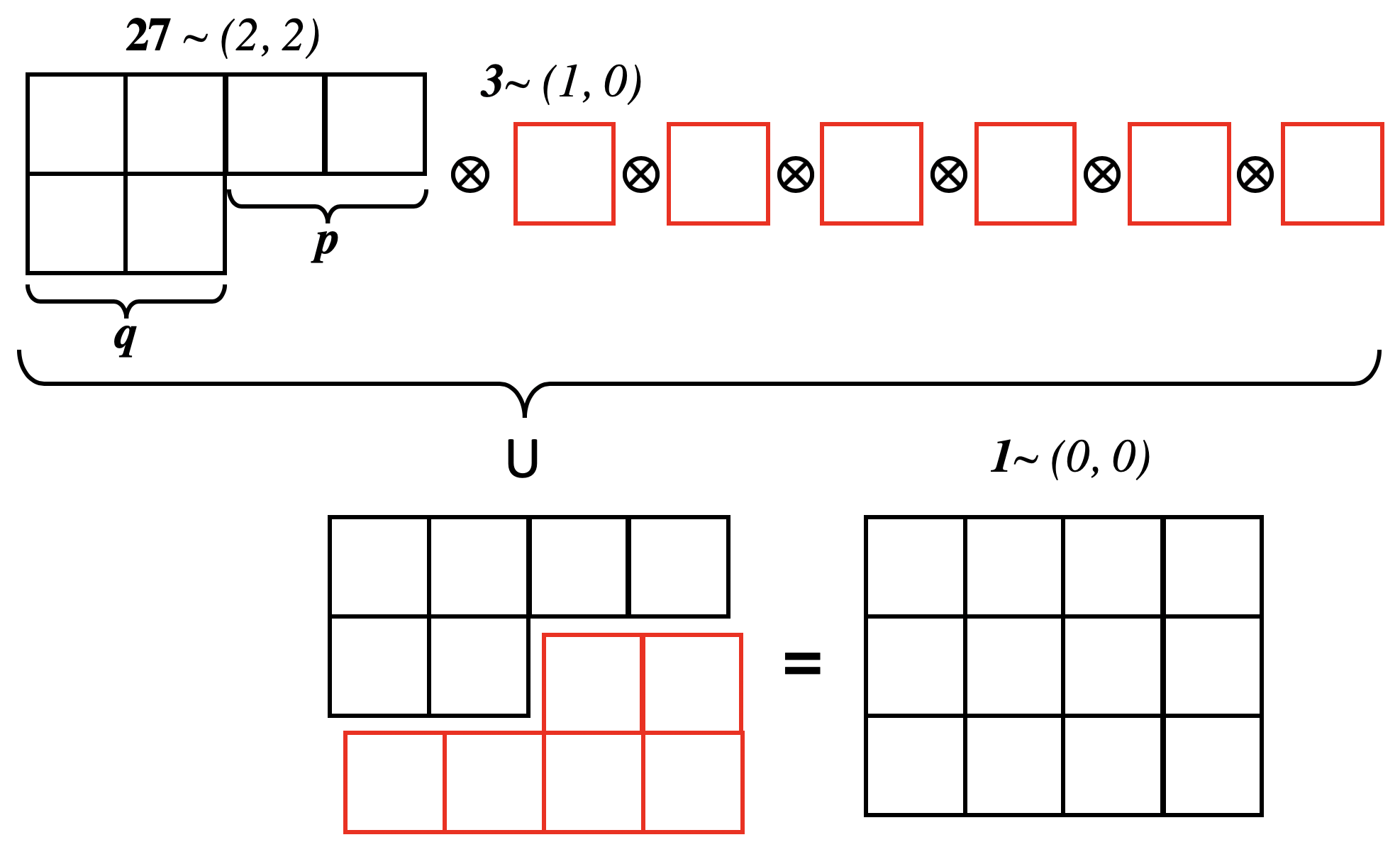}
    \caption{A schematic representation of how the minimal number of direct products of the fundamental representation (here, for $\boldsymbol{27}\sim(2,2)$, that number is $2p+q=6$) produces a Young tableau containing the trivial representation.}
    \label{fig:fund}
\end{figure}

Visually, one simply needs to fill the Young tableaux of the representation $(p,q)$ to a rectangle of $3\times(p+q)$ boxes; this requires $3p+3q-(2q+p)=2p+q$ additional boxes, or copies of the fundamental representation, as shown in fig.~\ref{fig:fund}.

This result is easily generalized, by the same argument, to SU($N$), where irreducible representations are labeled by $(p_1,p_2,...,p_{N-1})$, and the number of fundamental representations is given by
\begin{equation}
    k_N=p_{N-1}+2p_{N-2}+...+(N-2)p_2+(N-1)p_1.
\end{equation}

A more formal proof of the statement above can  be obtained from the Schur-Weyl duality\footnote{I am grateful to Martin Weissman for pointing this out to me.} \cite{etingof2009introduction}: the direct product of $k$ copies of the fundamental representation $\boldsymbol{N}$ of SU($N$) decomposes into a direct sum over of irreducible representations labeled by all ordered partitions $\lambda_1\ge \lambda_2...\ge \lambda_i$ of $k$ with $i\le N$. The question of whether, given a representation $X$, the representation $X\otimes \boldsymbol{N}^{\otimes k}$  contains the trivial representation is equivalent to asking whether $\boldsymbol{\overline{N}}$ is contained in the Schur-Weyl duality sum. But given that for a representation $X\sim(p_1,p_2,...,p_{N-1})$ the conjugate representation $\boldsymbol{\overline{X}}\sim(p_{N-1},p_{N-2},...p_2,p_1)$, whose Young tableaux contains exactly $k_N=p_{N-1}+2p_{N-2}+...+(N-2)p_2+(N-1)p_1$ boxes, the $\boldsymbol{\overline{X}}$ certainly belongs to the Schur-Weyl duality decomposition; this also proves that $k_N$ is the smallest possible number $k$ such that  $X\otimes \boldsymbol{N}^{\otimes k}$ contains the trivial representation, since $k_N-1$ would not have a sufficient number of Young tableaux to produce $\boldsymbol{\overline{X}}$ in the Schur-Weyl duality decomposition.
\newpage

\section{The minimal number of gluons, quarks, antiquarks}
\label{sec:minimal}

{ If a hypothetical, massive new strongly-interacting particle ${\bf X}\sim (p,q)$ existed, it would hadronize into a color-singlet hadron. Phenomenologically, it is of interest to understand the structure of the lowest-lying (``smallest'') hadronized state. To this end, while conclusive results can only be derived with non-perturbative techniques such as lattice simulations (see e.g. \cite{Foster:1998wu} and references therein), it is possible, and it has historically been the preferred route, to operate within the formalism of the MIT bag model \cite{DeGrand:1975cf,DeRujula:1975qlm,Rebbi:1975ns}. In this context, in addition to quarks and antiquarks (and their generalizations in SU($N$), the lowest-lying hadrons also include {\em valence gluons}, corresponding to ${\bf 8}\sim (1,1)$, or $(\underbrace{1,0,\dots,0,1}_{N-1})$ for SU($N$). 

In a {\em perturbative} context, the question above is equivalent to the question of which product of quark, antiquark and gluon fields lead to a gauge invariant operator connecting the new particle ${\bf X}$ with Standard Model fields. In the notation of Ref.~\cite{Kumar:2011tj}, one can write the relevant operators as
\begin{equation}
    {\cal O}_i^{(n)}=\frac{C_i^{(n)}}{\Lambda^{k_i}} X\ \widetilde{\cal O}_i^{(n)}(g,q,\overline{q}),
\end{equation}
where $\widetilde{\cal O}^{(n)}_i(g,q,\overline{q})$ is an operator containing exactly $n$ Standard Model quark, antiquark and/or gluon fields, $C_i^{(n)}$ is a dimensionless constant, and $\Lambda^{k_i}$ is the suppression scale of the operator $ {\cal O}_i^{(n)}$, with integer $k_i$  the mass dimension, which additionally depends on the Lorentz structure of the ${\bf X}$. Therefore, our results here are not limited to bound states, but also e.g. to the multi-jet phenomenology of possibly unstable ${\bf X}$ states produced at high-energy colliders \cite{Kumar:2011tj}.

It is convenient to define the notion of $N$-ality (triality in the case $N=3$) of an irreducible representation $(p_1,p_2,\dots,p_{N-1})$ as
\begin{equation}
t=\left(\sum_{j=1}^{N-1}\ j p_j\right) {\rm mod}\ N.
    \label{eq:nality}
\end{equation}
Notice that for SU(3), $t=(p+2q)$ mod 3. Any product of irreducible representations that contains the trivial representation must have $t=0$. This is the starting point to build the ``smallest'' SU($N$) hadrons: the minimal addition to the exotic $X\sim (p_1,p_2,\dots,p_{N-1})$ with $N$-ality $t$ is 
\begin{equation}
    Q_{N-t}\sim(n_1,n_2,\dots,n_{N-1}),\quad n_j=1\  {\rm for}\ j=N-t,\quad n_j=0\ \ {\rm otheriwise},
\end{equation}
since the $N$-ality of $Q_j$ is $j$. For instance, in SU(3) this means that representations with triality $t=2$ will need one additional ``quark'', i.e. $\mathbf{3}\sim (1,0)$ and those with $t=1$  one additional ``antiquark'', i.e. $\overline{\mathbf{3}}\sim (0,1)$.

Let us now calculate the result of ${\bf X}\otimes Q_{N-t}$. Start with the easiest case of $N=3$, and ${\bf X}\sim(p_1,p_2)$.  
If $t=1$, {we need to calculate 
\begin{eqnarray}
    (p_1,p_2)\otimes (0,1)&=&\ \ (p_1-1,p_2)\quad ({\rm iff}\ p_1\ge1) \label{eqrep1}  \\
                      \label{eqrep2}      &&+(p_1+1,p_2-1)\quad ({\rm iff} \ p_2\ge1)\\
                        \nonumber     &&+(p_1,p_2+1).
\end{eqnarray}
Of the representations on the right hand side of the equation above, the optimal one to achieve the goal of obtaining the trivial representation with the minimal possible number of products of the adjoint is $(p_1-1,p_2)$ if $p_1>0$ and $(1,p_2-1)$ if $p_1=0$, since those are the representations corresponding to (i) the smallest number of boxes in their Young diagrams, and (ii) the fewest boxes in the first row (this condition minimizes the number of copies of the adjoint representation, for diagrams with the same number of boxes, by avoiding adding additional columns to get to the trivial representation, making the representation in (\ref{eqrep1}), if possible, preferable to that in (\ref{eqrep2}) despite both having the same number of boxes). Similarly, if $t=2$, the optimal representation is $(p_1,p_2)\otimes (1,0)\supset(p_1,p_2-1)$ if $p_2>0$, and $(p_1-1,1)$ if $p_2=0$.}

Next, given a representation with null triality, say in the case of SU(3) ${\bf X}^\prime\sim(p,q)$, we ought to calculate the {\em minimal number} of copies of gluon fields  (or ``valence gluons'', in the language, again, of the MIT bag model) ${\bf 8}\sim (1,1)$ leading to an exotic colorless hadron. To this end, let me explicitly calculate the product 
\begin{eqnarray}
{\bf X}^\prime\otimes (1,1) & = & (p+1,q+1)\\
& \oplus & (p+2,q-1)\quad ({\rm iff}\ q\ge1)\\
& \oplus & (p-1,q+2)\quad ({\rm iff}\ p\ge1)\\
& \oplus & (p,q)\\
& \oplus & (p+1,q-2)\quad ({\rm iff}\ q\ge2)\label{eq4}\\
& \oplus & (p-2,q+1)\quad ({\rm iff}\ p\ge2)\label{eq5} \\
& \oplus & (p-3,q+3)\quad ({\rm iff}\ p\ge3)\\
& \oplus & (p-1,q-1)\quad ({\rm iff}\ p\ge1\ {\rm and}\ q\ge1). \label{eq7}
\end{eqnarray}
Assume now that $p=q$. In this case, from Eq~(\ref{eq7}) above, the minimal number of ``valence gluons'' required to obtain the trivial representation (0,0) is exactly $p$.

Assume instead that $p>q$ (the conjugate case $p<q$ follows immediately, see below). Since the triality of ${\bf X}^\prime$ is zero, $p=q+3k$, with $k$ a positive integer. First, ${\bf X}^\prime\otimes (1,1)^{\otimes q}$ contains, from Eq~(\ref{eq7}) above, representation $(3k,0)$. Then, from Eq.~(\ref{eq5}) above, $(3k,0)\otimes (1,1)^{\otimes k}$ contains representation $(k,k)$; finally, further direct product with $(1,1)^{\otimes k}$ will then contain the trivial representation (from Eq.(\ref{eq7} again). In total, one needs $q+2k=(2p+q)/3$ copies of the adjoint representation. Notice that this is the minimal number of such copies as well, since as shown in the previous section this corresponds to the same number of boxes of the minimal product of the fundamental representation that contains the trivial representation.

The demonstration for $p<q$ is identical: let $q=p+3k$; ${\bf X}^\prime\otimes (1,1)^{\otimes p}$ contains, from Eq~(\ref{eq4}) above, representation $(0,3k)$. Then, from Eq.~(\ref{eq4}), $(0,3k)\otimes (1,1)^{\otimes k}$ contains representation $(k,k)$; and, as above, further direct product with $(1,1)^{\otimes k}$ will then contain the trivial representation (from Eq.(\ref{eq7}) again). In total, this is $p+2k=(p+2q)/3$ which, as it should, is the symmetric version of what found above under $p\leftrightarrow q$.

In summary, for SU(3) I find that the number of copies $n_q$ of the fundamental representation ${\bf 3}\sim(1,0)$, of copies $n_{\bar q}$ of the antifundamental representation ${\bf \overline{3}}\sim(0,1)$, and of copies $n_g$ of the adjoint representation ${\bf 8}\sim(1,1)$ needed for the product $$(p,q)\otimes (0,1)^{\otimes n_{\bar q}}\otimes (1,0)^{\otimes n_q}\otimes (1,1)^{\otimes n_g}\supset (0,0)$$ 
is as follows (I also include the lowest dimension ``portal operator'' ${\cal O}_{\rm min}$ for each case, and its Lorentz structure for the minimal case; additional operators containing more quark, antiquark and gluon fields, and different Lorentz structures, can be produced in a straightforward manner by adding products of combinations containing the trivial representation, such as $\bar q q$, $qqq$, $\bar q\bar q\bar q$, $gg$ etc.):
\begin{itemize}
    \item $t=(p+2q){\rm mod\ 3}=0$, $n_q=n_{\bar q}=0$, and $n_g=(2p+q)/3$ if $p\ge q$, $n_g=(2q+p)/3$ if $p<q$; ${\cal O}_{\rm min}=C\frac{g^{n_g}X}{\Lambda^{n_g-3}}$ ($X$ scalar or vector)
    \item $t=1$, $n_q=0$, $n_{\bar q}=1$, and $n_g= (2(p-1)+q)/3$ if $p\ge q$, $n_g= (2(q-1)+p)/3$ if $p<q$; ${\cal O}_{\rm min}=C\frac{g^{n_g}(X\Gamma\bar q)}{\Lambda^{n_g-1}}$, ($X$ a Dirac fermion, $\Gamma$ a generic Dirac gamma matrix structure)
    \item $t=2$, $n_q=1$, $n_{\bar q}=0$, and  $n_g= (2p+q-1)/3$ if $p \ge q$, $n_g= (2q+p-1)/3$ if $p<q$; ${\cal O}_{\rm min}=C\frac{g^{n_g}(\bar X\Gamma q)}{\Lambda^{n_g-1}}$, ($X$ a Dirac fermion, $\Gamma$ a generic Dirac gamma matrix structure) 
\end{itemize}
\newpage
Let me now make a few non-trivial examples\footnote{Here and in what follows I acknowledge the use of the {\tt Susyno} package \cite{Fonseca:2011sy}.}:
\begin{itemize}
    \item {\bf 10}$\sim$(3,0), $t=0$, $n_q=0,\ n_{\bar q}=0$, $n_g=(2p+q)/3=2$:
$${\bf 10}\otimes{\bf 8}\otimes{\bf 8}={\bf 1}\oplus(4\times {\bf 8})\oplus(4\times {\bf 10})\oplus(2\times {\bf \overline{10}})\oplus(5\times {\bf 27})\oplus{\bf 28}\oplus(4\times {\bf {35}})\oplus {\bf \overline{ 35}}\oplus(2\times{\bf 64})\oplus{\bf 81}$$
while
$${\bf 10}\otimes{\bf 8}={\bf 8}\oplus{\bf 10}\oplus{\bf 27}\oplus{\bf 35} $$
${\cal O}_{\rm min}=\Lambda C g^2 X$, $X$ scalar or vector;
\end{itemize}
 \begin{itemize}   
    \item {\bf 15}$\sim$(2,1), $t=1$, $n_q=0,\ n_{\bar q}=1$, $n_g=(2(p-1)+q)/3=1$:
$${\bf 15}\otimes{\bf\overline{3}}\otimes{\bf 8}={\bf 1}\oplus(4\times {\bf 8})\oplus(3\times {\bf 10})\oplus(2\times {\bf \overline{10}})\oplus(4\times {\bf 27})\oplus(2\times {\bf 35})\oplus{\bf\overline{35}}\oplus{\bf 64}$$
${\cal O}_{\rm min}=Cg(X\Gamma\bar q)$, 
$X$ a Dirac fermion, $\Gamma$ a generic Dirac gamma matrix structure;
    \end{itemize}
    \begin{itemize}  
    \item {\bf 15$^\prime$}$\sim$(4,0), $t=1$, $n_q=0,\ n_{\bar q}=1$, $n_g=(2(p-1)+q)/3=2$:
    $${\bf 15}\otimes{\bf\overline{3}}\otimes{\bf 8}\otimes{\bf 8}={\bf 1}\oplus(6\times {\bf 8})\oplus(8\times {\bf 10})\oplus(3\times {\bf \overline{10}})\oplus(11\times {\bf 27})\oplus(5\times {\bf 28})\oplus(13\times {\bf 35})\oplus(3\times {\bf\overline{35}})$$
    $$\oplus(8\times{\bf 64})\oplus(2\times{\bf 80})\oplus(7\times{\bf 81})\oplus{\bf\overline{81}}\oplus(2\times{\bf 125})\oplus{\bf 154}$$
    while
    $${\bf 15}\otimes{\bf\overline{3}}\otimes{\bf 8}={\bf 8}\oplus(2\times {\bf 10})\oplus(2\times {\bf 27})\oplus{\bf 28}\oplus(3\times {\bf 35})\oplus{\bf 81}$$
    ${\cal O}_{\rm min}=C\frac{g^2(X\Gamma\bar q)}{\Lambda}$, 
$X$ a Dirac fermion, $\Gamma$ a generic Dirac gamma matrix structure;
\end{itemize}
    \begin{itemize}  
    \item ${\bf\overline{6}}\sim$(0,2), $t=2$, $n_q=0,\ n_{\bar q}=1$, $n_g=(2q+p-1)/3=1$:
    $$  {\bf \overline{6}}\otimes{\bf\overline{3}}\otimes{\bf 8}={\bf 1}\oplus(3\times {\bf 8})\oplus{\bf 10}\oplus(2\times {\bf \overline{10}})\oplus(2\times {\bf 27})\oplus {\bf\overline{35}}$$
    ${\cal O}_{\rm min}=Cg(X\Gamma\bar q)$, 
$X$ a Dirac fermion, $\Gamma$ a generic Dirac gamma matrix structure.

\end{itemize}

The generalization to SU($N$), $N>3$, is relatively straightforward, although care must be taken in handling cases where one or more of the $p_i=0$. Consider ${\bf X}=(p_1,p_2,\dots,p_{N-1})\otimes Q_j$, with $1\le j\le N-1$. If $p_{N-j}>0$, then $${\bf X}\otimes Q_j\supset (p_1,p_2,\dots,p_{N-j}-1,\dots,p_{N-1}),\quad p_{N-j}>0. $$
If $p_{N-j}=0$ but $p_{N-j\pm1}>0$, then $${\bf X}\otimes Q_j\supset(p_1,p_2,\dots,p_{N-j-1}-1,1,p_{N-j+1}-1\dots,p_{N-1}),\quad p_{N-j}=0,\ p_{N-j\pm1}>0 .$$ 
If $p_{N-j-1}=0$, then if $p_{N-j-2}>0$ $${\bf X}\otimes Q_j\supset(p_1,p_2,\dots,p_{N-j-2}-1,1,0,p_{N-j+1}-1\dots,p_{N-1}), $$ etc., until, if $p_1=0$, $${\bf X}\otimes Q_j\supset(1,0,\dots,0,p_{N-j+1}-1\dots,p_{N-1}),\quad p_i=0\ \forall i=1,\dots,N-j;\ p_{N-j+1}>1.$$ Similarly, if $p_{N-j+1}=0$, but $p_{N-j-1}>0$ and $p_{N-j+2}>0$
$$
{\bf X}\otimes Q_j\supset(p_1,p_2,\dots,p_{N-j-1}-1,0,1,p_{N-j+2}-1,\dots,p_{N-1}), 
$$
etc., until, if $p_{N-1}=0$
$$
{\bf X}\otimes Q_j\supset(p_1,p_2,\dots,p_{N-j-1}-1,0,\dots,0,1),\quad   p_i=0\ \forall i=j+1,\dots,N-1;\ p_{N-j-1}>1.
$$
Finally, of course if ${\bf X}=(\delta_{1,N-j},\delta_{2,N-j},\dots,\delta_{N-1,N-j})$, then evidently ${\bf X}\otimes Q_j\supset(0,\dots,0)$.

The generalization of the calculation of the number of ``valence gluons'' for $N>3$ is not straightforward, bar in a few special cases. After obtaining the representation ${\bf X}^\prime\equiv{\bf X}\otimes Q_j$ with vanishing $N$-ality  as outlined above, the algorithmic procedure to obtain a product of representations ${\bf X}^\prime\otimes G_N^{\otimes k}\supset {\bf 1}$, with $G_N=(1,0,\dots,0,1)$ the adjoint representation, containing the trivial representation ${\bf 1}$, is as follows: first note that the Young tableau for the adjoint representation is given by a ``doublet'' of boxes on the first row, with $N-2$ boxes below it; given that in a Young tableau the length of the rows are of decreasing length from the top, the optimal choice to locate the $N-2$ boxes and the doublet is to place the latter in the lowest possible row, and to place the additional single boxes on the lowest possible rows they can be placed at.

The smallest possibel number of copies of the adjoint representation needed so that the product ${\bf X}^\prime\otimes G_N^{\otimes k_{\rm min}}\supset {\bf 1}$ is 
$$
k_{\rm min}=\frac{1}{N}\sum_{j=1}^{N-1}(N-j)p_j,
$$
where ${\bf X}^\prime=(p_1,p_2,\dots,p_{N-1})$. One can verify that for representations with vanishing $N$-ality, $k_{\rm min}$ is an integer number that corresponds to the number of empty boxes after subtracting those filled by the Young tableaux of representation ${\bf X}^\prime$, in a rectangle of length $\sum_{j=1}^{N-1}p_j$ and height $N$, divided by $N$. In the generic case, additional columns, and thus additional copies of the adjoint representations, are needed to produce a direct product that contains the trivial representation. I leave the full solution of the $N>3$ case to future work.


}



With the results outlined above, assuming the electric charge $Q_X$ of a new hypothetical strongly-interacting particle $X$ belonging to a representation $\boldsymbol{X}\sim(p,q)$ is known, it is possible to calculate both the electric charge of the ``smallest'' hadron $Q_H$, and, generally, of any hadron containing $X$. Given the number of quarks $n_q$, antiquarks $n_{\overline{q}}$ and gluons $n_g$ listed in Tab.~\ref{tab:table}, the possible values of the charge of the smallest hadron $H$ are the following:
\begin{equation}
    -\frac{1}{3}n_q-\frac{2}{3}n_{\overline{q}}\le Q_H-Q_X\le \frac{2}{3}n_q+\frac{1}{3}n_{\overline{q}}.
\end{equation}
Any other hadron $H^\prime$ could only have electric charge $Q_{H^\prime}=Q_H+k$ for integer $k$.

Notice that all and only the representations with triality zero exclusively contain gluons in their ``smallest hadron''. Thus, it is only those representations that will yield hadronic bound states with integer charge if the ``new physics particle'' is neutral or of integer charge. I indicate those representations in green in Tab.~\ref{tab:table}. Notice that this set of representations includes {\em all} real (self-adjoint) representations ($p,p$). 

\begin{table}[]
    \centering
     {\tiny
     \begin{tabular}{|c|c|c|c|c|c|c|c|}
     
     \hline
          $p$	&	$q$	&	dim&		$t$	& $T(R)$ &	$n_g$		& $n_{\bar q}$	&	$n_q$	\\
          \hline
0	&	0	&	{\bf\color{gr}	1	}	&	0 & 0	&	0	&	0	&	0	\\
1	&	0	&	{\bf	3	}	&	1 &3 	&	0	&	1	&	0	\\
0	&	1	&	{${\bf\overline{	3	}}$} &3 	&	2	&	0	&	0	&	1	\\
2	&	0	&	{\bf	6	}	&	2	& 15 &	1	&	0	&	1	\\
0	&	2	&	{${\bf\overline{	6	}}$} 	&	1 &15	&	1	&	1	&	0	\\
1	&	1	&	{\bf\color{gr}	8	}	&	0	&	18&1	&	0	&	0	\\
3	&	0	&	{\bf\color{gr}	10	}	&	0	&	45&2	&	0	&	0	\\
0	&	3	&	{\color{gr}${\bf\overline{	10	}}$}	&	0 &45	&	2	&	0	&	0	\\
4	&	0	&	{\bf	15$^\prime$	}	&	1	&105&	2	&	1	&	0	\\
2	&	1	&	{\bf	15	}	&	1	&60&	1	&	1	&	0	\\
1	&	2	&	{${\bf\overline{	15	}}$}	&	2 &	60 &	1	&	0	&	1	\\
0	&	4	&	{${\bf\overline{	15	}}^\prime$}	&	2 &105	&	2	&	0	&	1	\\
5	&	0	&	{\bf	21	}	&	2	& 210&	3	&	0	&	1	\\
0	&	5	&	{${\bf\overline{	21	}}$}	&	1 & 210	&	3	&	1	&	0	\\
3	&	1	&	{\bf	24	}	&	2 &150&		2	&	0	&	1	\\
1	&	3	&	{${\bf\overline{	24	}}$}	&	1& 150	&	2	&	1	&	0	\\
2	&	2	&	{\color{gr}\bf	27	}	&	0 &162	&	2	&	0	&	0	\\
6	&	0	&	{\color{gr}\bf	28	}	&	0	&378&	4	&	0	&	0	\\
0	&	6	&	{\color{gr}${\bf\overline{	28	}}$}	&	0&378	&	4	&	0	&	0	\\
4	&	1	&	{\color{gr}\bf	35	}	&	0&315	&	3	&	0	&	0	\\
1	&	4	&	{\color{gr}${\bf\overline{	35	}}$}	&	0&315	&	3	&	0	&	0	\\
7	&	0	&	{\bf	36	}	&	1&630	&	4	&	1	&	0	\\
0	&	7	&	{${\bf\overline{	36	}}$}	&	2&630	&	4	&	0	&	1	\\
3	&	2	&	{\bf	42	}	&	1&357	&	2	&	1	&	0	\\
2	&	3	&	{${\bf\overline{	42	}}$}	&	2&357	&	2	&	0	&	1	\\
8	&	0	&	{\bf	45	}	&	2&990	&	5	&	0	&	1	\\
0	&	8	&	{${\bf\overline{	45	}}$}	&	1&990	&	5	&	1	&	0	\\
5	&	1	&	{\bf	48	}	&	1&588	&	3	&	1	&	0	\\
1	&	5	&	{${\bf\overline{	48	}}$}	&	2&588	&	3	&	0	&	1	\\
9	&	0	&	{\color{gr}\bf	55	}	&	0&1485	&	6	&	0	&	0	\\
0	&	9	&	{\color{gr}${\bf\overline{	55	}}$}	&	0&1485	&	6	&	0	&	0	\\
4	&	2	&	{\bf	60	}	&	2&690&		3	&	0	&	1	\\
2	&	4	&	{${\bf\overline{	60	}}$}	&	1&690	&	3	&	1	&	0	\\
6	&	1	&	{\bf	63	}	&	2&1008	&	4	&	0	&	1	\\
1	&	6	&	{${\bf\overline{	63	}}$}	&	1&1008	&	4	&	1	&	0	\\
3	&	3	&	{\color{gr}\bf	64	}	&	0&720	&	3	&	0	&	0	\\
10	&	0	&	{\bf	66	}	&	1	&2145&	6	&	1	&	0	\\
0	&	10	&	{${\bf\overline{	66	}}$}	&	2 & 2145&		6	&	0	&	1	\\
11	&	0	&	{\bf	78	}	&	2&3003	&	7	&	0	&	1	\\
0	&	11	&	{${\bf\overline{	78	}}$}	&	1&3003	&	7	&	1	&	0	\\
7	&	1	&	{\color{gr}\bf	80	}	&	0&1620	&	5	&	0	&	0	\\
1	&	7	&	{\color{gr}${\bf\overline{	80	}}$}	&	0&1620	&	5	&	0	&	0	\\
5	&	2	&	{\color{gr}\bf	81	}	&	0 &1215	&	4	&	0	&	0	\\
2	&	5	&	{\color{gr}${\bf\overline{	81	}}$}	&	0& 1215&		4	&	0	&	0	\\
4	&	3	&	{\bf	90	}	&	1&1305	&	3	&	1	&	0	\\
3	&	4	&	{${\bf\overline{	90	}}$}	&	2&1305	&	3	&	0	&	1	\\
12	&	0	&	{\color{gr}\bf	91	}	&	0&4095	&	8	&	0	&	0	\\
0	&	12	&	{\color{gr}${\bf\overline{	91	}}$}	&	0&4095	&	8	&	0	&	0	\\
8	&	1	&	{\bf	99	}	&	1&2475	&	5	&	1	&	0	\\
1	&	8	&	{${\bf\overline{	99	}}$}	&	2&2475	&	5	&	0	&	1	\\
\hline
     \end{tabular}
}

    \caption{\small List of all irreducible representations of SU(3) with dimension smaller than 100, the corresponding triality $t=2q+p$, index of the representation $T(R)$ with the normalization convention of Eq.~(\ref{eq:normconv}) (the normalization convention in e.g. Ref.~\cite{Slansky:1981yr} would give $T(R)/3$), with the minimal number of gluons, antiquarks and quarks needed to form a color-singlet hadron. The smallest hadrons for representations in green only contain gluons, and, if the ``new physics particle'' belonging to that representation is electrically neutral, would also be electrically neutral.}
    \label{tab:table}
\end{table}

\subsection{Constraints from Asymptotic Freedom}\label{sec:asfreedom}
{Generally, a theory with matter fermions or scalars in a large-dimensional representation will not be asymptotically free  \cite{Eichten:1982pn}. Since here I assume the new particle $X$ to be very heavy, with a mass much larger than the typical binding energy $E_b$ of the $X$ bound states with quarks and gluons discussed here, since $E_b\approx \Lambda_{\rm QCD}$, this is not a concern: the $X$ would effectively not contribute to the beta function of QCD at those scales, in just the same way that only light quark flavors, and not heavy quark flavors, contribute to the running of the QCD coupling below $\Lambda_{\rm QCD}$ (this fact is sometimes referred to as the Appelquist-Carazzone decoupling theorem \cite{Appelquist:1974tg}). Even at very high energies of order the mass of the $X$, the issue of jeopardizing asymptotic freedom for QCD is somewhat mute, since at those high energies the QCD gauge group might be embdedded in a larger gauge group, with additional gauge bosons that would alter any conclusion based on additional matter fields only. We discuss a few examples of embeddings of the $X$ in grand unification setups below.

The caveats above notwithstanding, and while as explained above this would not prevent the formation of the bound states discussed here, it is worthwhile to briefly summarize the implications of the additional requirement that the theory, including the new $X$ particle, be asymptotically free. The weakest constraints arise if the SM is solely augmented by a single new real scalar $X_s$ in a representation $R(X_s)$. Notice that this is only possible if the representation is real. The requirement of asymptotic freedom for SU(3), after including all $N_f=6$ SM quark flavors, which presumably are all lighter than the mass scale of the new particle $X$, can be expressed as 
\begin{equation}\label{eq:asfree}
11\times T({\rm adj})-4\times N_f\times T(R_f)-\frac{1}{2}T(R(X_s))\ge0,
\end{equation}
where $T(R)$ is the trace normalization factor for  representation $R$, and 
\begin{equation}
T(R)=\frac{C_2(R)d(R)}{d(G)},
\end{equation}
with $d(G)$ the dimension of the adjoint representation ($d(G)=N^2-1$ for SU($N$)), and $C_2(R)$ the quadratic Casimir operator of the representation, which for $N=3$ and representation $(p,q)$, and with the normalization convention \cite{White:1992aa}
\begin{equation}\label{eq:normconv}
2NX_R^aX_R^a=C_2(R){\mathds{1}},
\end{equation} reads
\begin{equation}
C_2((p,q))=6p+2p^2+6q+2q^2+2pq.
\end{equation} 
Notice that with this normalization convention $T({\rm adj})=C_2(1,1)=18$, $T(R_f)=3$, and the asymptotic freedom condition in Eq.~(\ref{eq:asfree}), with $N_f=3$, becomes
\begin{equation}
    T(R(X_s))\le252.
\end{equation}

This limits the possible representations that would leave SU(3) asymptotically free above the $m_X$ scale to the following real representations (see the $T(R)$ values, with the normalization convention adopted here, associated with all SU(3) representations of dimension less than 100 in the fifth column of tab.~\ref{tab:table}):
\begin{equation}
    {\rm real\ scalar:}\ {\bf 1},\ {\bf 8},\ {\bf 27}. 
\end{equation} 
For a complex scalar\footnote{Note that the use of complex scalar fields is not directly connected with the properties of the representations involved.  After all, one can always take a complex field and by doubling the components represent them as real fields (I thank Howard Haber for bringing this to my attention).}, $T(R(X_s))\le126$, giving the following possibilities:
\begin{equation}
    {\rm complex\ scalar:}\ {\bf 1},\ {\bf 3},\ {\bf\overline{3}},\  {\bf 6},\ {\bf\overline{6}},\ {\bf 8},\ {\bf 10},\ {\bf\overline{10}},\  {\bf 15},\ {\bf\overline{15}},\  {\bf 15^\prime},\ {\bf\overline{15}^\prime}. 
\end{equation} 
For a Weyl or Majorana fermion, the condition becomes $T(R(X_W))\le63$, , giving the following possibilities:
\begin{equation}
    {\rm Weyl\ or \ Majorana\ fermion:}\ {\bf 1},\ {\bf 3},\ {\bf\overline{3}},\  {\bf 6},\ {\bf\overline{6}},\ {\bf 8},\ {\bf 10},\ {\bf\overline{10}},\  {\bf 15},\ {\bf\overline{15}}.
\end{equation} 
Finally, for a Dirac fermion\footnote{As explained in Ref.~\cite{Dreiner:2008tw}, given a collection of two-component fermions grouped into a sum of multiplets that transform irreducibly under SU(3), if a multiplet transforms under a real representation of SU(3) then the corresponding fermion mass eigenstates are Majorana fermions, while if a multiplet transforms under a complex representation, then the corresponding mass eigenstates are Dirac fermions (I again thank Howard Haber for reminding me of this fact). I therefore do not list real representations for the Dirac fermion case.}, $T(R(X_D))\le31$, , giving the following possibilities:
\begin{equation}
    {\rm Dirac\ fermion:}\ {\bf 3},\ {\bf\overline{3}},\  {\bf 6},\ {\bf\overline{6}}.
\end{equation} 

As mentioned above, the asymptotic freedom constraints for SU(3) are relaxed when considering embeddings of Standard Model gauge interactions in a larger grand unification (GUT) gauge group, since the additional gauge fields tend to further stabilize the gauge coupling(s). Notice that the requirement that a GUT be asymptotically free is, however,  not limited to gauge couplings, but also to Yukawa couplings and scalar quartic couplings, which we do not consider hereafter (in any case, the asymptotic freedom of gauge couplings is a {\em necessary} condition). Finally, I also note that, as mentioned above, the extrapolation of asymptotic freedom to GUT scales may not be required due to the Appelquist-Carazzone theorem \cite{Appelquist:1974tg}; nonetheless, there might be exceptions to this, including speculation that asymptotic freedom is necessary for any consistent field theory \cite{Fradkin:1975rz, Fradkin:1975yt}.

I consider here three grand unification setups, for illustrations:
\begin{enumerate}
    \item Georgi-Glashow SU(5) \cite{Georgi:1974sy}: the matter fermions Weyl fields are embedded in (three copies of) the representations ${\bf \overline{5}}$ and {\bf 10}, the up and down Higgses in a {\bf 5} and ${\bf \overline{5}}$, and the real ``GUT scalars'' that cause the breaking of SU(5) in the adjoint representation {\bf 24} \cite{Georgi:1974sy}.
    \item For SO(10) I assume the matter fermions are contained in (three copies of) the {\bf 16}, the Higgses in the {\bf 10}, and the ``GUT scalars'' in the {\bf 45} \cite{Baez:2009dj}
    \item for E$_6$ matter fields are in (three copies of) the {\bf 27}, and I consider (following Ref.~\cite{Mohapatra:1985xm}) two possible symmetry breaking patterns with the following scalar content: (i): {\bf 650}, 2$\times${\bf 78}, {\bf 27} and {\bf 351}, and (ii) 2$\times${\bf 27}, {\bf 351}.
\end{enumerate}
The requirement of asymptotic freedom for any of the theories listed above is
\begin{equation}
    11\times T({\rm adj})-2\times 3 \times T(R_f) - \sum_i N_{S_i}\times T(R_{S_i})=q^{\rm GUT}\ge 0
\end{equation}
where $R_f$ identifies the representation to which the Standard Model (Weyl) matter fermions belong, and $S_i$ the (complex) scalars in the theory (real scalars contributing half that). The additional $X$ particle under consideration here would need to fit in an additional GUT multiplet with dimension at least as large as the dimension $R_X$ to which the $X$ belongs; the requirement of asymptotic freedom for the GUT under consideration is then $T(R^{\rm GUT}_X)\le q^{\rm GUT}$

I find the following results:
\begin{eqnarray}
    q^{SU(5)}&=&11\times 10-2\times3\times1-2\times3\times3-2\times1-10/2=79,\\
        q^{SO(10)}&=&11\times 8-2\times3\times2-1-8/2=71,\\
            q^{E_6(ii)}&=&11\times 4-2\times3\times1-50/2-4/2\times 2-1 =8,\\
            q^{E_6(ii)}&=&11\times 4-2\times3\times1-25-2\times 1=11.\\
\end{eqnarray}
 As a result, from Tab~\ref{tab:gut}, I find that the maximal dimension for the SU(3) representation to which the $X$ belongs is:
\begin{enumerate}
    \item SU(5): 75 for a real scalar (the $X$ would be accommodated in the real representation {\bf 75} of SU(5)), 70 for a complex scalar (the $X$ would be accommodated in the complex representation {\bf 70} of SU(5), etc.), 50 for a Weyl fermion, and 15 for a Dirac fermion;
    \item SO(10): 320 for a real scalar, 210 for a complex scalar, 144 for a Weyl fermion, and 16 for a Dirac fermion (the {\bf 45} or {\bf 54} would imply Majorana mass eigenstates since they are real representations);
    \item E$_6$ (i) and (ii): 78 for a real scalar or a Weyl fermion, 27 for a complex scalar or a Dirac fermion.
\end{enumerate}

\begin{table}[]
    \centering
    \begin{tabular}{||c|c||c|c||c|c||}
    \hline
       \ \ \ \ \ \ \  SU(5)\ \ \ \ \ \ \    & $T(R)/T({\rm fund})$ &\ \ \ \ \ SO(10)\ \ \ \ \   & $T(R)/T({\rm fund})$ &\ \ \ \ \ \ \ \ E$_6$\ \ \ \ \ \ \ \ & $T(R)/T({\rm fund})$ \\
       \hline
       {\bf 5} & 1 & {\bf 10}* & 1 & {\bf 27} & 1 \\
       {\bf 10} & 3 & {\bf 16} & 2 & {\bf 78}* & 4 \\
       {\bf 15} & 7 & {\bf 45}* & 8 & {\bf 351} & 25 \\
       {\bf 24}* & 10 & {\bf 54}* & 12 & {\bf 351$^\prime$} & 28 \\
       {\bf 35} & 28 & {\bf 120}* & 28 & {\bf 650}* & 50 \\
       {\bf 40} & 22 & {\bf 126} & 35 & & \\
       {\bf 45} & 24 & {\bf 144} & 34 & & \\
       {\bf 50} & 35 & {\bf 210}* & 56 & & \\
       {\bf 70} & 49 & {\bf 210$^\prime$}*& 77 & & \\
       {\bf 70$^\prime$} & 84 & {\bf 320}* & 96 & & \\  
       {\bf 75}*  & 50 & {\bf 560} & 182 & & \\
       {\bf 105}  & 91 & & & & \\
       {\bf 126}  & 105 & & & & \\
       {\bf 126$^\prime$}  & 210 & & & & \\
       {\bf 160}  & 168 & & & & \\
       {\bf 175}  & 140 & & & & \\
       \hline
\end{tabular}
    \caption{Lowest-dimensional non-trivial representations for SU(5), SO(10) and E$_6$, and the corresponding values for $T(R)/T({\rm fund})$. Real representations are indicated with a *.}
    \label{tab:gut}
\end{table}
     }

\section{Conclusions}
\label{sec:conclusions}
I proved that the smallest number of copies $k$ of the fundamental representation (1,0) of SU(3) such that the direct product of irreducible representation ${\bf X}\sim(p,q) \otimes (1,0)^{\otimes k}$ contains the trivial representation $(0,0)$ is $k=2p+q$; I generalized this result to SU($N$), where for irreducible representation $(p_1,p_2,...,p_{N-1})$, $
    k_N=p_{N-1}+2p_{N-2}+...+(N-2)p_2+(N-1)p_1$. {I outlined the structure of the smallest-possible product of representations containing quark, antiquark, and gluon fields as well the ${\bf X}$, which corresponds to the smallest bound-state hadron, or to the minimal QCD gauge-invariant operator connecting the new strongly interacting particle to Standard Model fields. I gave exact results for $N=3$ and outlined the $N>3$ case. A corollary of these results is the calculation of the electric charge of the resulting bound states. Finally, I discussed constraints stemming from demanding  asymptotic freedom both in the case of pure QCD, and in the case of a few example embeddings of the new particle in a grand unification setup.}

\section*{Acknowledgements}
I gratefully acknowledge helpful conversations with Robert Boltije and Martin Weissman, and Daniel Davis and especially Howard Haber for feedback and suggestions on this manuscript. I also acknowledge discussions with Hiren Patel and Wolfgang Altmannshofer. { I am very grateful to the anonymous Referee for important and critical suggestions on an earlier version of this manuscript, and to Tesla Jeltema for help with some of the results presented in Sec.~\ref{sec:minimal}.} This work is partly supported by the U.S.\ Department of Energy grant number de-sc0010107.

\bibliography{hadrons}
\end{document}